# First-Principles Calculations of Born Effective Charges and Spontaneous Polarization of Ferroelectric Bismuth Titanate

**Amritendu Roy [1], Rajendra Prasad [2], Sushil Auluck [2] and Ashish Garg [1]**

[1] Department of Materials and Metallurgical Engineering

[2] Department of Physics

Indian Institute of Technology Kanpur; Kanpur- 208016; INDIA

**CORRESPONDING AUTHOR**

Ashish Garg

Email- ashishg@iitk.ac.in

Tel- +91-512-2597094

Fax- +91-512-2597505

**ABSTRACT**

In this study, we present the results of our first-principles calculations of the band structure, density of states and the Born effective charge tensors for the ferroelectric (ground state *B1a1*) and paraelectric (*I4/mmm*) phases of bismuth titanate. The calculations are done using the generalized gradient approximation (GGA) as well as the local density approximation (LDA) of the density functional theory. In contrast to the literature, our calculations on *B1a1* structure using GGA and LDA yield smaller indirect band gaps as compared to the direct band gaps, in agreement with the experimental data. The density of states shows considerable hybridization among Ti $3d$, Bi $6p$ and O $2p$ states indicating covalent nature of the bonds leading to the ferroelectric instability. The Born effective charge tensors of the constituent ions for the ground state (*B1a1*) and paraelectric (*I4/mmm*) structures were calculated using the Berry phase method. This is followed by the calculation of the spontaneous polarization for the ferroelectric *B1a1* phase using the Born effective charge tensors of the individual ions. The calculated value for the spontaneous polarization of ferroelectric bismuth titanate using different Born effective charges was found to be in the range of $55\pm13$ μC/cm$^2$ in comparison to the reported experimental value of ($50\pm10$ μC/cm$^2$) for single crystals. The origin of ferroelectricity is attributed to the relatively large displacements of those oxygen ions in the TiO$_6$ octahedra that lie along the $a$-axis of the bismuth titanate crystal.

**PACS Number (s):** 77.22.Ej, 85.50.Gk, 77.84.-s, 77.80.-e, 71.15.Mb

**Key words** – Bismuth Titanate, Ferroelectricity, Born effective charge,

Spontaneous Polarization

## I.     INTRODUCTION

Bismuth titanate ($Bi_4Ti_3O_{12}$ or, BiT), a ferroelectric perovskite oxide, has received tremendous attention, especially in thin film form, due to its promise in nonvolatile ferroelectric random access memory (NvFRAM) applications (1). BiT has a reasonably large spontaneous polarization ($P_s$), a high Curie temperature (948 K), low processing temperature (< 973 K) in thin film form, lead (Pb) free nature, fatigue resistance with lanthanide doping on Pt electrodes and low permittivity.  These attributes make this material suitable for various applications *e.g.* memories, piezoelectric devices and electro-optic devices (1-4). BiT belongs to the Aurivillius family of phases, represented by the general formula $Bi_2A_{x-1}B_xO_{3x+3}$ or $(Bi_2O_2)^{2+}(A_{x-1}B_xO_{3x+1})^{2-}$, where A is a mono-, di- or tri-valent cation such as $Na^+$, $Sr^{2+}$, $Bi^{3+}$ etc, B is a transition metal ion such as $Ti^{4+}$, $Ta^{5+}$, $Nb^{5+}$ etc and $x$ varies between 2 and 4. Its unit cell consists of alternate stacking of $(Bi_2O_2)^{2+}$ and perovskite like $(A_{x-1}B_xO_{3x+1})^{2-}$ layers arranged along the *c*-axis. BiT undergoes a displacive type of ferroelectric to paraelectric transition across the Curie temperature at $T_c \sim 948$ K (5). The high temperature paraelectric phase has a tetragonal crystal structure with space group *I4/mmm*.

Until recently, there has been considerable ambiguity regarding the crystal structure of the low temperature ferroelectric phase since experimental studies reported both orthorhombic and monoclinic structures. For instance, initial X-ray diffraction (XRD) studies performed by Aurivillius (6) indicated an orthorhombic structure with space group *Fmmm*. Later Dorrian *et al.* (7) showed the structure of the ferroelectric phase to be orthorhombic with space group *B2cb*, also supported by neutron diffraction studies conducted by Hervoches *et al.* (8). However, on the basis of XRD data, Dorrian *et al.* (7) obtained a non-zero polarization (~ 4 μC/cm$^2$) along *c*-axis which is quite unlikely for an orthorhombic *B2cb* structure since in this structure *b*-glide plane has its mirror plane perpendicular to *c*-axis and as a result, projections of all polarization vectors along *c*-axis should cancel out. On the other hand, structure refinement of the electron diffraction data of a BiT single crystal by Rae *et al.* (9) predicted a monoclinic *B1a1* structure. The absence of *b*-glide plane in monoclinic *B1a1* structure could explain the existence of small remnant polarization along the *c*-axis of the ferroelectric BiT. Temperature dependent Raman spectroscopy studies also showed a soft mode behavior at 150-200 K indicating a monoclinic distortion of the orthorhombic structure (10, 11). Though there are a large number of experimental works on the structural and electrical characterization of pure and doped BiT, the ambiguity on the equilibrium structure remained and needed theoretical studies.

First in-depth structure optimization of BiT was performed by Cai *et al.* (12) on the *I4/mmm* structure while electronic structure calculations based on the same structure were performed by Shimakawa *et al* (2) using optimized basis set with effective core potentials. Further investigations by Cai *et al.* (13) using the full-potential linear augmented plane wave (FP-LAPW) plus local orbital method within density functional theory (DFT) revealed that the ground state structure (ferroelectric phase) of BiT to be monoclinic (*B1a1*). The exchange and correlation effects were treated within generalized gradient approximation (GGA) using Perdew-Burke-Ernzerhof (PBE) gradient correction correlation functional (12, 13). The ground state *B1a1* structure of BiT was further substantiated by Shrinagar *et al.* (14), Perez-Mato *et al.* (15) and Hashimoto *et al.* (16). Studies by Cai *et al.* (12, 13) indicate that the ferroelectric response originates largely from the Ti $3d$-O $2p$ hybridization which is further enhanced by Bi $6s$, $6p$-O$2p$ hybridization.

Electrical characterization of BiT ceramics, single crystals and thin films show a wide range of spontaneous polarization ($P_s$) from 1 to 50 $\mu$C/cm$^2$ (17-25). Such a wide variation could be attributed to the sample quality, orientation and defect chemistry. To the best of our knowledge, in the only theoretical work on the polarization calculations of BiT carried out so far, Hashimoto *et al.* (16) calculated the spontaneous polarization of BiT using Berry phase method within the Simulation Tool for Atom Technology (STATE) which is based on the DFT. Though the calculated value was close to the experimental results, no accounts of contributions of individual ions to the overall polarization were provided. Moreover, calculations of Born effective charges (BECs) by Hashimoto *et al.* (26) were based on *B2cb* structure which is not the correct ground state structure. Since *B2cb* belongs to orthorhombic structure, the BEC tensors of individual ions would have three different values of the elements of the leading diagonal. Hashimoto *et al.* (16) ignored this and probably reported an average value. In the same way, they reported average values of BECs of ions in *I4/mmm* structure which should actually consist of two or three different values in the leading diagonal.

It has been shown earlier by us and other coworkers that *B1a1* structure is the ground state structure of BiT (14-16). Here, we report the results of our first-principles calculation of Born effective charge (BEC) tensors and spontaneous polarization on this structure using Berry phase method (27). Moreover by comparing BECs and polarizations in the *B1a1* structure with the results of Hashimoto *et al*. (16), we could ascertain the effect of structure on polarization. We have also calculated BEC tensors of each ion in the high temperature *I4/mmm* structure. Our electronic structure calculations and polarization calculations are compared with earlier theoretical calculations and experimental data. In this paper, first we discuss the methodologies of our calculations in section II, followed by presentation and analysis of results of our calculations in section III.

## II.    CALCULATION DETAILS

In the present work, the lattice parameters and the initial co-ordinates of the atoms in the *B1a1* structure were taken from our earlier work (14) where the cell parameters were relaxed to zero force and zero pressure conditions using the GGA. The lattice parameters are $a = 5.4289 \overset{\circ}{A}$, $b = 5.4077 \overset{\circ}{A}$, $c = 32.8762 \overset{\circ}{A}$, and $\beta = 90.08^{\circ}$. The volume of the unit cell is 964.85 $Å^3$. Our entire calculations were carried out in the framework of first-principles density functional theory (28). Vienna ab-initio simulation package (VASP) (29, 30) was used with projector augmented wave method (PAW) (31). The Kohn-Sham equation (32, 33) was solved using the exchange correlation function of Perdew and Wang (34,

35) for generalized gradient approximation (GGA), and the local density approximation (LDA) schemes. We included five valence electrons for Bi ($6s^2p^3$), 4 for Ti ($3d^34s^1$) and 6 for O ($2s^22p^4$). In our earlier work (14), we optimized the ground state structure of bismuth titanate with cut-off energy of 400 eV which yielded good convergence and hence, was also used for present calculations. Conjugate gradient algorithm (36) was used for the structural optimization. All the calculations were performed at 0 K.

For electronic structure calculation, Monkhorst-Pack (37) 8×8×8 mesh was used. Born effective charge and spontaneous polarization were calculated using Berry phase method (27) with Monkhorst-Pack (37) 3×3×1 and 4×4×4 mesh. We started our calculations with a 4×4×4 mesh. However, to reduce the computational time, we reduced the mesh size to 3×3×1 mesh and yielded Born effective charge values sufficiently close to those calculated using a 4×4×4 mesh. Therefore all calculations were performed with 3×3×1 mesh. Similar calculations have also been performed for the paraelectric *I4/mmm* structure for which the initial lattice parameters (38) are taken as $a = b = 3.8524 \,\overset{\circ}{\mathrm{A}}$ and $c = 33.197 \,\overset{\circ}{\mathrm{A}}$. The volume of this unit cell is 492.68 Å$^3$. The unit cell parameters were relaxed to zero force and zero pressure conditions keeping the shape of the cell unaltered. Fig 1(a) and 1(b) show the unit cells of BiT in its paraelectric (*I4/mmm*) and ferroelectric (*B1a1*) forms respectively. The number of atoms in the tetragonal unit cell is half the number of atoms in the *B1a1* unit cell. Upon cooling, across the Curie temperature, the paraelectric phase transforms to the ferroelectric phase. In order to calculate the displacement vectors (vector defining relative displacement of a particular ion while transforming from the paraelectric to the ferroelectric phase), the paraelectric *I4/mmm* cell was transformed to an equivalent tetragonal cell having correspondence with the ferroelectric *B1a1* cell. This transformation is schematically shown in Fig. 2. The lattice parameters "*a*" and "*b*" and ionic coordinates were transformed in a way shown in equation (1).

$$
\left.
\begin{aligned}
&a_{\mathrm{mod}} = b_{\mathrm{mod}} = \sqrt{2}\,a_{orig} \ \text{ and } \ c_{\mathrm{mod}} = c_{orig} \\
&(0,0,z)_{T,orig} \longleftrightarrow (0,0,z),(0.5,0.5,z)\,|_{T,\mathrm{mod}} \\
&(0.5,0.5,z)_{T,orig} \longleftrightarrow (0,0.5,z),(0.5,0,z)\,|_{T,\mathrm{mod}} \\
&(0,0,0)_{T,orig} \longleftrightarrow (0,0,0),(0.5,0.5,0)\,|_{T,\mathrm{mod}} \\
&(0,0.5,0)_{T,orig} \longleftrightarrow (0.25,0.75,0),(0.75,0.25,0)\,|_{T,\mathrm{mod}} \\
&(0.5,0,0)_{T,orig} \longleftrightarrow (0.25,0.25,0),(0.75,0.75,0)\,|_{T,\mathrm{mod}} \\
&(0,0.5,z)_{T,orig} \longleftrightarrow (0.25,0.75,z),(0.75,0.25,z)\,|_{T,\mathrm{mod}} \\
&(0.5,0,z)_{T,orig} \longleftrightarrow (0.25,0.25,z),(0.75,0.75,z)\,|_{T,\mathrm{mod}}
\end{aligned}
\right\} \qquad (1)
$$

where $a_{\mathrm{mod}}$, $b_{\mathrm{mod}}$, $c_{\mathrm{mod}}$ and $a_{orig}$, $c_{orig}$ are the lattice vectors in the derived tetragonal and original *I4/mmm* cells respectively. Fig. 2 and eq. (1) also show that the transformation of *I4/mmm* cell only takes place on the *a-b* plane and the *c*-axis lattice parameter remains unchanged.

The Born effective charge (BEC) tensor $Z^{*}_{K,\gamma\alpha}$ of atom $k$ can be linked either to the change in polarization $P_{\gamma}$ induced by the periodic displacement $\tau_{k,\alpha}$ or to the force $F_{k,\alpha}$ induced on atom $k$ by an electric field $\xi_{\gamma}$.

$$Z^*_{k,\gamma\alpha} = V \frac{\delta P_\gamma}{\delta \tau_{k,\alpha}} \qquad (2)$$

$$= \frac{\delta F_{k,\alpha}}{\delta \xi_\gamma}$$

$$= -\frac{\partial^2 E}{\partial \xi_\gamma \partial \tau_{k,\alpha}} \qquad (3)$$

Hence, it appears as a mixed second derivative of the total energy $E$ (per unit cell volume).

Spontaneous polarization ($P_s$) can be described as the difference between the polarization of a ferroelectric phase and high temperature paraelectric phase. In the present study, we have calculated polarization difference ($\Delta P$) between the ferroelectric (*B1a1*) and the paraelectric (*I4/mmm*) phases assuming that a continuous adiabatic transformation takes place that includes scaling the internal strain with a parameter $\lambda$ such that $0 \leq \lambda \leq u$, where, '$u$' is the ionic displacement calculated considering the coordinates of each ion in the ferroelectric *B1a1* phase and in the paraelectric, *I4/mmm* phase. The polarization in the paraelectric phase is assumed to be zero, P[0], since there is no dipole. On the other hand polarization in the ferroelectric phase will have a finite value, P[$u$] and this polarization arises continuously over the transformation temperature from state P[0] to state P[$u$] as described by the following equation

$$\int_0^u Z^*(u)du = [\text{P}(u)\text{-P}(0)] = u\text{Z}(u) \qquad (4)$$

where

$$Z(u) = \frac{1}{u} \int_0^u Z^*(u) du \qquad\qquad (5)$$

Eq. (5) defines Z (*u*) as the mean value of Z*(*u*) from 0 to *u*.

## III.        RESULTS AND DISCUSSION

### (A) Electronic Structure: Band Structure and Density of states

In the present study, all calculations are based on optimized *I4/mmm* structured paraelectric and optimized *B1a1* structured ferroelectric phases. Structural relaxation of either of the two phases (*I4/mmm* and *B1a1*) are performed using the GGA and the LDA. Structural parameters of the relaxed cells are reported in Table 1. The GGA band structure, along high symmetry directions for BiT in the paraelectric (*I4/mmm*) and ferroelectric (*B1a1*) phases are shown in Fig 3(a) and 3(b), respectively. In Fig. 3(b) the band structure is denser than that in Fig. 3(a) since the number of atoms in the unit cell is doubled. From the band structure shown in Fig 3 (a), it is seen that the material in its paraelectric phase, possesses an indirect band gap ($E_{indir}$) of 1.53 eV along $M \rightarrow \Gamma$ and a direct band gap ($E_{dir}$) of 2.22 eV at the Brillouin-zone center, $\Gamma$, indicating the semiconducting nature of the phase. The LDA band structure (not shown here) also shows almost similar characteristics with corresponding indirect and direct band gaps as 1.63 eV and 2.29 eV, respectively. The indirect and direct band gaps in the *B1a1* phase, as shown in Fig 3 (b), are 2.17 eV ($E \rightarrow \Gamma$) and 2.30 ($\Gamma \rightarrow \Gamma$) eV. Subsequent LDA calculation on *B1a1* structure gives an indirect band gap of 2.07 eV ($E \rightarrow \Gamma$) and a direct band gap of 2.25 eV ($\Gamma \rightarrow \Gamma$). These gaps are also compared with the previous calculations in Table 2. We find that both the direct and indirect band gaps calculated using the GGA, in the paraelectric (*I4/mmm*) structure show good agreement with the work of Shimakawa *et al*. (2) who employed plane wave (PW) method. Also, it is interesting to note that our calculations on *I4/mmm* and *B1a1* structure yield $E_{indir} < E_{dir}$ which is in agreement with the available experimental data for the *B1a1* phase.

In contrast, a previous study (13) using FP-LAPW showed $E_{indir} > E_{dir}$ for the *B1a1* structure, as seen from Table 2. This difference is attributed to the difference in the methods of calculations. For instance, Cai *et al*. (13) used full potential method whereas the present work is based on pseudopotential method. The type of potential significantly affects the calculations. Results also depend upon whether exchange correlation effects are treated within GGA or LDA. Being the simplest approximation, LDA assumes that the exchange correlation energy at every point in the system is identical to that of a uniform electron gas of the same density. Though LDA has been used for many systems, it suffers from the tendency of overbinding and as a result, it frequently underestimates the lattice parameters while overestimating bulk moduli, phonon frequencies, and cohesive energies. GGA, on the other hand, corrects the LDA exchange correlation energy with a gradient function of charge density and is, in principle, expected to provide a better representation for open structures such as BiT. Finally, calculated results also depend upon the type of exchange correlation functional and the software used for the calculation. For example, Cai *et al*. (13) used gradient-corrected correlation functional of Perdew– Burke–Ernzerhof (PBE) using Wien2k code while we have used correlation functional of Perdew and Wang and used VASP to solve Kohn-Sham equation.

Another study by Machado *et al.* (39) using FLAPW-LO with exchange and correlation effects treated within LDA showed an indirect gap of ~1.2 eV and a direct gap of ~ 2.0 eV for the paraelectric *I4/mmm* phase. These authors also relaxed the structure to the experimental volume as against our approach of relaxing the structure to zero force and zero pressure. It is interesting to note that the calculated gaps by Machdo *et al.* (39) are much smaller than the experimental gaps in Table 2. This is a well known problem with the LDA and GGA which underestimate the band gaps since their exchange correlations do not include the effect of excited states (40-43).

The total density of states (DOS) of the two structures, *B1a1* and *I4/mmm*, calculated using the GGA are shown in Fig 4. Upon comparing with the band structure shown in Fig 3, we note that the bands for paraelectric phase in this energy range show a downward shift leading to smaller band gap for this phase. We can further divide the band structure and DOS into four parts. From the partial DOS we can determine the angular momentum character of various sections. The lowest energy bands, (-19 to -16 eV) mainly consist of O 2*s* and Ti 3*d* states while Bi 6*s* states are located in the energy range -11 to -9 eV. The uppermost part of the valence band (VB) consists of contributions from Bi 6*p*, Ti 3*d* and O 2*p* states. In this energy window, Ti 3*d* state is mainly built by Ti 2 and Ti 2′ ions while the contribution from Ti 1 ion is quite small. Hybridization of Bi 6*s*, Ti 3*d* and O 2*p* states was also observed in this section of the DOS. Above the Fermi level, the conduction band (CB) is dominated by Ti $3d$ states with minor contributions from Bi $6p$ and O $2p$ states. Also, there is considerable hybridization between Ti $3d$, Bi $6p$ and O $2p$ in the CB as observed from the figure. Fig. 4 also shows that the shape of the DOS plots of *I4/mmm* and *B1a1* are broadly similar with minor changes in the location and shape of the structures. For example above the Fermi level, the conduction band starts at ~ 1.5 eV for *I4/mmm* structure whereas it begins at ~2.2 eV for *B1a1* structure. Our results using the VASP code are in agreement with the results of Cai *et al.* (13) who used the FP-LAPW method. We also calculated the density of states using the LDA (not shown here) and found no major difference with the GGA DOS.

**(B) Born Effective Charges (BECs)**

Born effective charge (BEC or, $Z^*$), also known as transverse or dynamic effective charge, is a fundamental quantity that manifests coupling between lattice displacements and electrostatic fields. Advances in ab-initio techniques now enable one to determine BEC theoretically using perturbation theory or finite difference in polarization. It has been found that BEC values are relatively insensitive to isotropic volume change but are strongly affected by changes in atom positions associated with the phase transitions. BEC is important in terms of theoretical study of ferroelectrics since the ferroelectric transition takes place from the competition of long range coulomb interactions and short range forces. Thus long range coulomb interactions are responsible for the splitting of frequencies of LO (longitudinal optical) and TO (transverse optical) phonons. BEC is an indicator of long range coulomb interactions for this splitting. First-principles calculations depict that $Z*$ is anomalously large, sometimes twice the static ionic charge. Why some charges are so large has remained a fundamental problem thus far. A band by band decomposition of $Z*$ is a useful tool to investigate the role of covalency and ionicity without any preliminary hypothesis on the orbital that interact.

In the present work, we have evaluated the BEC tensors of each ion in ferroelectric (*B1a1*) and paraelectric (*I4/mmm*) structures of BiT by slightly displacing each ion, one at a time, along three directions of Cartesian co-ordinates and calculating the resulting difference in polarization, using Berry phase method (27). Since the *B1a1* structure has very little monoclinic distortion compared to the orthorhombic *B2cb* structure, we have assumed the BEC tensor will be characteristic to the lattice vectors (parallel to the right handed Cartesian co-ordinate axes) of the bismuth titanate unit cell. We have calculated the BECs of the constituent ions for *B1a1* and *I4/mmm* structures using both the GGA and LDA and found that both the results follow the same trend with some differences in the BEC values. Table 3 and Table 4 present the BEC tensor of each ion in *B1a1* and *I4/mmm* structures of bismuth titanate calculated using the GGA.

First, we consider the BECs of ferroelectric *B1a1* structure shown in Table 3. The formal valence of Bi, Ti and O in $Bi_4Ti_3O_{12}$ are +3, +4 and -2 respectively. In the present structure, constituent ions developed a maximum effective charge of 5.36 for Bi, 7.23for Ti, and -6.15 for O showing +79 %, +81 % and -207 % change from the static charge respectively for each ion . Due to site symmetry, the BEC tensor of each ion has anisotropic diagonal elements with presence of finite off-diagonal elements as well. In comparison to the GGA results, our calculation of BECs using the LDA shows that the maximum effective charge of 5.41 for Bi, 7.27 for Ti and -5.30 for O showing +80 %, +82% and -165% change from the static charge respectively for each ion. For comparison with the literature, we have used GGA results as these results are believed to be more accurate than LDA results. In contrast to our results, Hashimoto *et al.* (16) reported a maximum charge of 4.72 for Bi, 6.07 for Ti and -3.58 for O with +57%, +51% and -79% change from the static charge in the ferroelectric *B2cb* structure. These BECs are significantly different from the BECs obtained in the present work and the difference could be a manifestation of the effect of structure on the BECs since the two calculations were performed on two different structures. However, the bond-valence calculation of effective charges of each ion in monoclinic *B1a1* BiT by Shulman *et al.* (44) and Withers *et al.* (45) show a considerably smaller charges of ions compared to that of the present calculation. This is because of the difference in techniques as bond-valence method is not ab-initio in nature(45) and therefore prone to giving different results when compared to ab-initio methods such as ours.

We also evaluated the BECs of ions in the paraelectric *I4/mmm* structure. The maximum changes in the charges of different ions are: +81 % for Bi, +62 % for Ti and -176 % for O respectively when calculated using the GGA. BECs of constituent ions calculated using the LDA results, however, differ significantly from the GGA results and in this case the maximum change in effective charges are: +99% for Bi, +84% for Ti and -226 % for O ions respectively. In contrast, corresponding values reported by Hashimoto *et al.* (16) were +95% for Bi, +86% for Ti and -116% for O. We see from Table 4 that the BEC tensors of O1 and O5 do not maintain symmetry unlike other ions of the structure. This is not an anomaly as eq. (3) suggests that BEC is a mixed second derivative of the total energy, with respect to the electric field and atomic displacement. Previous literature on tetragonal $KNbO_3$ also reports such asymmetric BEC tensors of O ions (46). Having a closed shell like nature, the charge carried by an ion would be close to its nominal charge. On the other hand, if covalent character of the bonds is present in the compound, a significant amount of charge flows through it when ions are displaced from their parent co-ordinates (47, 48).

Table 3 shows that the BECs of ions in $Bi_4Ti_3O_{12}$ are significantly higher than their nominal charges and one can, to a first approximation, conclude that there is a significant amount of covalent character present in the structure. Large BECs also point towards the relative displacements of neighboring ions against each other giving rise to large polarization. Large BECs of the cations compared to their nominal charges also indicate that both Bi and Ti act as donors while O ions act as acceptors. This is also manifested by the smaller BECs of O ions. Presence of off-diagonal charges in Ti and O ions confirms that the bonds among 6 O ions and the central Ti ion in the $TiO_6$ octahedra are of covalent nature. Large magnitude of charges associated with the Bi ions also indicates about the possible hybridization of Bi and neighboring O ions.

A comparison of the present GGA calculated BECs of the constituent ions in BiT with other perovskites shows that while Bi in the ferroelectric (*R3c*)

BiFeO$_3$ develops 82% increase in the effective charge (49), corresponding increase in the *B1a1* structure of BiT is 79%. Such a large anomalous charge is attributed to the strong hybridization between Bi *6p* and O *2p* bonds, also demonstrated by Ravindran *et al.* (49) in BiFeO$_3$. In the present study, we find that Ti in ferroelectric BiT has a maximum of 81% increase in the effective charge over its static charge. There is similar increase in Ti effective charge in other works such as 79% increase in case of cubic BaTiO$_3$, 78% increase in case of cubic SrTiO$_3$, 76.5% increase in case of cubic PbTiO$_3$ (48). Oxygen, on the other hand, finds a maximum of 207% reduction in the static charge in the *B1a1* structure of BiT. In comparison to this, the effective charge of O ion decreases by 185% in case of cubic BaTiO$_3$, 183% in case of cubic SrTiO$_3$, 250% in case of cubic KNbO$_3$, 192% in case of cubic PbTiO$_3$, 176% in case of tetragonal PbTiO$_3$ (48) and 226% in case of cubic PbMg$_{1/3}$Nb$_{2/3}$O$_3$ (50). We find that Born effective charge is quite sensitive to ion positions and crystal symmetry since the effective charge for a particular ion varies significantly in *B1a1* and *I4/mmm* structures and the three O ions have different charges in the same phase. This observation is in agreement with a previous study by Choudhury *et al.* (51) on PbMg$_{1/3}$Nb$_{2/3}$O$_3$.

**(C) Spontaneous Polarization (P$_s$)**

Spontaneous polarization (P$_s$) is defined as the difference between the polarizations of a ferroelectric phase and a high temperature paraelectric phase. Heating beyond Curie temperature results in transformation of monoclinic (*B1a1*) bismuth titanate into a non-ferroelectric tetragonal (*I4/mmm*) phase. The direction of P$_s$ in BiT single crystal lies in the monoclinic *a-c* plane and the magnitude of the component of $P_s$ along the *a*-axis of the monoclinic unit cell was first determined by Cummins *et al.* (24) yielding $P_s$ values of ~50±10 μC/cm$^2$ at 25°C. These authors also showed a small but non-zero $P_s$ along c-axis of 4±0.1 μC/cm$^2$ indicating a large anisotropy of $P_s$ (24). Therefore, the true spontaneous polarization in BiT has a magnitude of ~50 μC/cm$^2$ and the direction of $P_s$ vector makes an angle 4.5° with the major surface (approx. normal to *c*-direction) (24). Variations in reported data have been found for different samples. The reported anomaly in polarization values can be attributed to the quality of sample, sample orientation during measurement and defect chemistry.

Based upon our GGA calculation, in the present work we have calculated spontaneous polarization along three directions, *x*, *y* and *z*, assuming that Z(*u*) values are identical to the Z*(*u*) of the ferroelectric phase. The displacement vector '***u***' was calculated considering the coordinate of each ion in the ferroelectric *B1a1* phase and that in the paraelectric, *I4/mmm* phase. Lattice parameters and ionic coordinates for the *I4/mmm* phase were taken from the literature (43) and relaxed further to zero pressure and zero force conditions. In order to calculate the displacement vector (***u***), the tetragonal cell was transformed to another tetragonal cell (compatible to the *B1a1* monoclinic cell) as shown schematically in Fig 2 and as explained in section II. Using Z$^*$ values from Table 3 and calculating displacement vectors between ferroelectric monoclinic and paraelectric tetragonal cell, we calculated the $P_s$ for the ferroelectric BiT using eq. (4). The magnitude of three components of the spontaneous polarization vectors are $P_x = 39.39$ μC/cm$^2$, $P_y = 15.87$ μC/cm$^2$ and $P_z = 5.59$ μC/cm$^2$. The magnitude of the $P_s$ was calculated to be $P_s = 42.83$ $\mu$C/cm$^2$ which is well within the range of the experimental value (24) for a good quality sample. The polarization vector lies at 21.94° to the *a*-axis on the *a-b* plane. The polarization vector also makes an angle of 7.49° to the *a*-axis on the '*a*-c' plane. The deviation from the experimental results could be due to the fact that while reported experimental data were recorded at the room temperature, our calculations were done at 0 K.

In order to estimate the possible error in the theoretical value of $P_s$, we calculated $P_s$ of BiT in three different ways. Firstly, as we have explained in the preceding paragraph, we calculated $P_s$ using the $Z^*(u)$ of the ferroelectric (*B1a1*) phase. Secondly, we calculated the value of $P_s$ using the calculated $Z^*(u)$ of the paraelectric phase, as given in Table 4. This value was ~ 67.08 $\mu$C/cm$^2$ which is larger than experimental values (24). It was found in an earlier study (49) that the Born charges for higher symmetry structure give larger polarization of the ferroelectric phase. With $Z^*(u)$ of the paraelectric phase, taken from literature (16), the calculated $P_s$ of the *B1a1* structure is ~66.94 $\mu$C/cm$^2$. Finally, we have also calculated the $P_s$ using average values of BECs between *B1a1* and *I4/mmm* structures (Tables 3 and 4). This method yielded a $P_s$ of 54.94 $\mu$C/cm$^2$. Previously, Ravindran *et al.* (49) also calculated $P_s$ in BiFeO$_3$ using similar methodology and reported agreement between theoretical and experimental values of $P_s$. Thus the calculated values of $P_s$ using the GGA fall in the range of 43-68 $\mu$C/cm$^2$.

We have also calculated $P_s$ of ferroelectric BiT using the LDA. The $P_s$ calculated on the basis of our LDA calculation of the BECs in the *B1a1* structure is 55.84 $\mu C/cm^2$ whereas it is ~68.26 $\mu C/cm^2$ for the *I4/mmm* structure and these values agree reasonably well with our GGA results. A comparison of the $P_s$ calculated using different BECs under different schemes (LDA and GGA) is presented in Table 5.

Our results show that the spontaneous polarization of BiT calculated using the GGA and LDA with different BECs of the ferroelectric *B1a1* and paraelectric *I4/mmm* phases fall in the range ~ 55±13 $\mu C/cm^2$, overlapping with the experimental data. On the contrary, calculation of the spontaneous polarization using the static charges of the ions in the *B1a1* structure yield $P_s$ ~ 31.9 $\mu C/cm^2$ which is significantly less than the values calculated using the BECs. Moreover, the experimental values are closer to the value calculated using BECs. This again explains the presence of noticeable amount of covalency in the material because BECs of different ions are significantly larger than the static charges of the respective ions.

The relation between partial and total polarization is similar to the relation between partial and total DOS. Just as the partial DOS gives information regarding the relative contribution of the various atoms to the total DOS (besides the angular momentum characteristics), partial polarization provides information regarding the relative contribution of individual atoms to the total polarization. We have, therefore, calculated the partial polarization using Z* (in *B1a1* using GGA) and *u* (based upon GGA calculation) for individual element one at a time.

Partial polarization calculations show that the contributions to the total $P_s$ from the constituent ions along $x$-direction are, $P_{Bi,x} = -1.62 \ \mu C/cm^2$, $P_{Ti,x} = 27.60 \ \mu C/cm^2$ and $P_{O,x} = -65.37 \ \mu C/cm^2$, respectively. The negative sign, in case of O ions, is due to the negative sign of the BECs associated with the O ions. From the above results, it is evident that Bi ions contribute little to the overall polarization as most of the contributions arise from Ti and O ions. This is quite different from the case of multiferroic BiFeO$_3$ where most of the polarization originates from Bi ions (49). The displacement vectors along the $a$-axis, calculated as mentioned above, are plotted in Fig 5. It is observed that the relative displacements for the Bi ions in the ferroelectric phase while originating from its parent tetragonal phase are minimum. However, the displacements for Ti and O ions are almost one order of magnitude larger than those of the Bi ions. Therefore, the TiO$_6$ octahedra in ferroelectric phase are substantially distorted as compared to their original shape in paraelectric phase. This distortion leads to the separation of charge centers in the TiO$_6$ octahedra giving rise to ferroelectricity in bismuth titanate.O1, O3 and O6 ions within the TiO$_6$ polyhedron are displaced farthest along $a$-axis during the transformation from *I4/mmm* to *B1a1*. This leads to larger charge separation along $a$-axis and thus explains the origin of ferroelectricity along $a$-axis in *B1a1* structure.

**IV CONCLUSIONS**

We have performed electronic structure calculations for the paraelectric (*I4/mmm*) and ferroelectric (*B1a1*) phases of bismuth titanate using first-principles density functional theory with the exchange correlation function of Perdew and Wang for the GGA and LDA schemes as implemented in VASP. The GGA band structure and density of states of ferroelectric and paraelectric states of bismuth titanate indicate that the ferroelectric phase has an indirect band gap of 2.17 eV and the paraelectric phase has an indirect gap of 1.53 eV. The corresponding direct band gaps are 2.30 eV and 2.22 eV. The LDA values are not very different with a maximum difference of 0.1 eV. It is interesting to note that our calculation on *B1a1* structure yields $E_{indir} < E_{dir}$ which is in agreement with experimental data. The GGA density of states of *B1a1* structure show considerable hybridization among Ti 3d, Bi 6p and O 2p indicating the covalent nature of the bonds which would lead to the ferroelectric instability. The GGA Born effective charge calculations in the *B1a1* structure using Berry phase method show that the constituent ions develop a maximum change in effective charge of +79% for Bi, +81% for Ti and -207% for O with respect to the static charge. In contrast such changes in the paraelectric *I4/mmm* structure are +81% for Bi, +62% for Ti and -176% for O. The magnitude of the spontaneous polarization calculated from Born effective charges in the *B1a1* structure  was ~43 $\mu C/cm^2$ for the GGA and ~56 $\mu C/cm^2$ for the LDA which are in agreement with the experimental value of 50±10 $\mu C/cm^2$. According to the GGA calculation, the polarization vector lies almost on the *a-c* plane and makes a small angle (7.49°) to the *a*-axis. The polarization vector also makes an angle of 21.94° to the *a*-axis on the '*a-b*' plane. Spontaneous polarization calculations based on Born effective charges of high symmetry paraelectric (*I4/mmm*) structure and average Born effective charges of *B1a1* and *I4/mmm* structures showed polarization value to be within a range of $55 \pm 13 \mu C/cm^2$. Partial polarization calculations showed that Ti and O ions contribute most to the overall polarization. Smaller contribution from Bi ions is attributed to relatively short displacement of the Bi ions in comparison to the other ions. O1, O3 and O6 ions within the $TiO_6$ polyhedron are displaced farthest along *a*-axis during the transformation and this is the primary reason for polarization along *a*-axis of BiT.

**Acknowledgements**

Authors thank Dr. Umesh Waghmare, Dr. Roy Benedek and Dr. R. K. Kotnala for helpful discussions. AG acknowledges the financial support by Department of Science and Technology (India) through Ramanna Fellowship.

Table 1 Structural parameters of the relaxed unit cells of *B1a1* and *I4/mmm*

phases calculated using GGA and LDA along with literature data.

| Parameter | GGA | | | LDA | |
|---|---|---|---|---|---|
| | **B1a1** | **I4/mmm** [38] | **I4/mmm** | **B1a1** | **I4/mmm** |
| Lattice parameters: | | | | | |
| $a$ (Å) | 5.4289 | 3.8524 | 3.8082 | 5.3149 | 3.7356 |
| $b$ (Å) | 5.4059 | 3.8524 | 3.8082 | 5.2924 | 3.7356 |
| $c$ (Å) | 32.8762 | 33.197 | 32.8163 | 32.1860 | 32.1906 |
| Cell Volume (Å$^3$) | 964.85 | 492.68 | 475.92 | 905.35 | 449.21 |

Table 2– Band gaps for ferroelectric *B1a1* and paraelectric *I4/mmm* structures

| Method | Paraelectric (*I4/mmm*) phase | | Ferroelectric (*B1a1*) phase | |
|---|---|---|---|---|
| | Indirect band gap (eV) | Direct band gap (eV) | Indirect band gap (eV) | Direct band gap (eV) |
| **Present Study (GGA)** | **1.53** | **2.22** | **2.17** | **2.30** |
| **(LDA)** | **1.63** | **2.29** | **2.07** | **2.25** |
| PW (2) | 1.38 | 2.10 | - | - |
| FLAPW-LO (12, 13) | 1.87 | 2.75 | 2.65 | 2.57 |
| GGA (14) | - | - | 2.16 | - |
| FLAPW-LO (39) | 1.2 | 2.0 | - | - |
| Experiment (9) | - | - | - | 3.6 |
| Experiment (52) | - | - | 3.27 | 3.60 |

Table 3- Born effective charge tensors (BECs) in ferroelectric bismuth titanate ($B1a1$) structure using GGA. Primed (′) positions are equivalent positions to the unprimed ones generated due to the point group ($m$) operation.

| Ion | Z* (e) | | | | | | | | |
|------|---------|----------|----------|----------|----------|----------|----------|----------|----------|
| | $Z_{xx}$ | $Z_{xy}$ | $Z_{xz}$ | $Z_{yx}$ | $Z_{yy}$ | $Z_{yz}$ | $Z_{zx}$ | $Z_{zy}$ | $Z_{zz}$ |
| Bi 1 | 5.10 | -0.44 | 0.04 | -0.42 | 5.35 | -0.26 | -0.17 | 0.01 | 5.15 |
| Bi 1′ | 4.77 | 0.00 | -0.32 | -0.48 | 5.36 | 0.43 | -0.28 | -0.01 | 4.81 |
| Bi 2 | 4.83 | -18 | 0.36 | -0.17 | 5.12 | 0.76 | 0.29 | 0.33 | 4.81 |
| Bi 2′ | 4.38 | 0.15 | 0.28 | -0.37 | 5.07 | 1.31 | -0.15 | 0.35 | 3.81 |
| Ti 1 | 6.60 | -0.24 | -0.33 | -0.45 | 7.23 | 1.07 | -0.71 | -0.30 | 5.81 |
| Ti 2 | 5.84 | -0.39 | 0.28 | -0.37 | 6.39 | 1.00 | 0.07 | 0.27 | 6.32 |
| Ti 2′ | 5.84 | -0.39 | 0.28 | -0.11 | 6.17 | 0.40 | 0.15 | 0.27 | 6.22 |
| O 1 | -3.77 | -1.93 | -0.13 | -1.82 | -6.15 | 0.01 | 0.16 | 0.21 | -2.22 |
| O 1′ | -2.77 | 1.68 | -0.45 | 1.28 | -4.58 | 1.19 | -0.03 | 0.98 | -2.84 |
| O 2 | -3.13 | 0.22 | -0.07 | -0.50 | -3.14 | 0.72 | -0.12 | 0.21 | -2.70 |
| O 2′ | -3.12 | -0.06 | 0.09 | 0.12 | -3.12 | 0.09 | 0.00 | 0.22 | -2.74 |
| O 3 | -2.93 | -0.44 | -0.42 | -0.63 | -3.12 | 0.49 | -0.58 | 0.86 | -4.58 |
| O 3′ | -2.94 | 0.43 | 0.49 | -0.10 | -3.06 | 1.19 | -0.01 | 0.18 | -3.66 |
| O 4 | -2.09 | 0.20 | -0.25 | -0.36 | -2.33 | -0.50 | -0.33 | -1.20 | -4.65 |
| O 4′ | -1.97 | -0.08 | 0.15 | -0.58 | -2.30 | -0.61 | -0.15 | -0.93 | -4.29 |
| O 5 | -3.47 | -0.99 | -1.22 | 0.55 | -3.02 | -0.43 | -0.59 | -0.29 | -2.37 |
| O 5′ | -3.39 | 0.91 | -0.15 | 0.30 | -3.81 | 0.37 | 0.29 | -0.35 | -2.17 |
| O 6 | -3.62 | 1.79 | 0.25 | 1.88 | -4.30 | 0.66 | 0.22 | 0.74 | -2.34 |
| O 6′ | -2.95 | -1.60 | -0.86 | -1.38 | -4.31 | 0.11 | -0.15 | 0.35 | -2.31 |

Table 4- Born effective charge tensors (BECs) in paraelectric bismuth titanate (*I4/mmm*) structure using GGA.

| Ion | Z* (e) | | | | | | | | |
|------|--------|--------|--------|--------|--------|--------|--------|--------|--------|
|      | $Z_{xx}$ | $Z_{xy}$ | $Z_{xz}$ | $Z_{yx}$ | $Z_{yy}$ | $Z_{yz}$ | $Z_{zx}$ | $Z_{zy}$ | $Z_{zz}$ |
| Bi 1 | 5.43 | 0.00 | -0.01 | -0.01 | 5.42 | -0.01 | 0.00 | 0.00 | 4.47 |
| Bi 2 | 4.92 | 0.00 | -0.01 | 0.00 | 4.92 | -0.01 | 0.00 | 0.00 | 3.80 |
| Ti 1 | 6.47 | 0.00 | 0.00 | 0.00 | 6.47 | 0.00 | 0.00 | 0.00 | 5.53 |
| Ti 2 | 5.50 | 0.00 | 0.12 | 0.00 | 5.50 | 0.12 | 0.00 | 0.00 | 5.98 |
| O 1 | -2.25 | 0.00 | 0.00 | 0.00 | -5.51 | 0.00 | 0.00 | 0.00 | -2.49 |
| O 2 | -3.12 | 0.00 | -0.04 | 0.00 | -3.12 | 0.04 | 0.00 | 0.00 | -2.78 |
| O 3 | -3.09 | 0.00 | 0.03 | 0.00 | -3.09 | 0.03 | 0.00 | 0.00 | -4.48 |
| O 4 | -2.23 | 0.00 | 0.05 | 0.00 | -2.23 | 0.05 | 0.00 | 0.00 | -4.49 |
| O 5 | -4.63 | 0.00 | -0.09 | 0.00 | -2.60 | 0.03 | 0.00 | 0.00 | -2.23 |

Table 5 – Spontaneous Polarization ($P_s$) of ferroelectric (*B1a1*) bismuth titanate evaluated from the Born effective charges of BiT using LDA and GGA.

| Method | Polarization (µC/cm$^2$) | | | $P_s$ (µC/cm$^2$) |
|--------|--------|--------|--------|--------|
|        | $P_x$ | $P_y$ | $P_z$ | |
| GGA (BEC $_{B1a1}$) | 39.39 | 15.87 | 5.59 | 42.83 |
| GGA (BEC $_{I4/mmm}$) | 63.04 | 21.54 | 7.83 | 67.08 |
| LDA (BEC $_{B1a1}$) | 54.20 | 12.26 | 5.40 | 55.84 |
| LDA (BEC $_{I4/mmm}$) | 64.95 | 19.53 | 7.66 | 68.26 |
| GGA ( Avg. BEC$_{B1a1 \& I4/mmm}$) | 51.21 | 18.71 | 6.71 | 54.94 |
| LDA ( Avg. BEC$_{B1a1 \& I4/mmm}$) | 59.58 | 15.90 | 6.53 | 62.01 |

**Figure Captions**

Fig. 1- Unit cells of (a) the ferroelectric monoclinic *B1a1* cell. The volume of the relaxed unit cell is 964.85 $Å^3$. (b) paraelectric tetragonal (*I4/mmm*) cell. The volume of the relaxed unit cell is 475.92 $Å^3$.

Fig. 2- Transformation of paraelectric tetragonal (*I4/mmm*) cell to another tetragonal cell which is compatible to the monoclinic *B1a1* cell.

Fig. 3- Electronic band structure of (a) paraelectric bismuth titanate with *I4/mmm* structure (b) ferroelectctric bismuth titanate with *B1a1* structure, calculated using the GGA.

Fig. 4- Density of states of ferroelectric *B1a1* structure and paraelectric *I4/mmm* structure calculated using the GGA. Contributions arising from different elements are also indicated.

Fig. 5- Displacement vectors of individual ions along *a*- axis when paraelectric structure is transformed to the ferroelectric structure across the Curie temperature

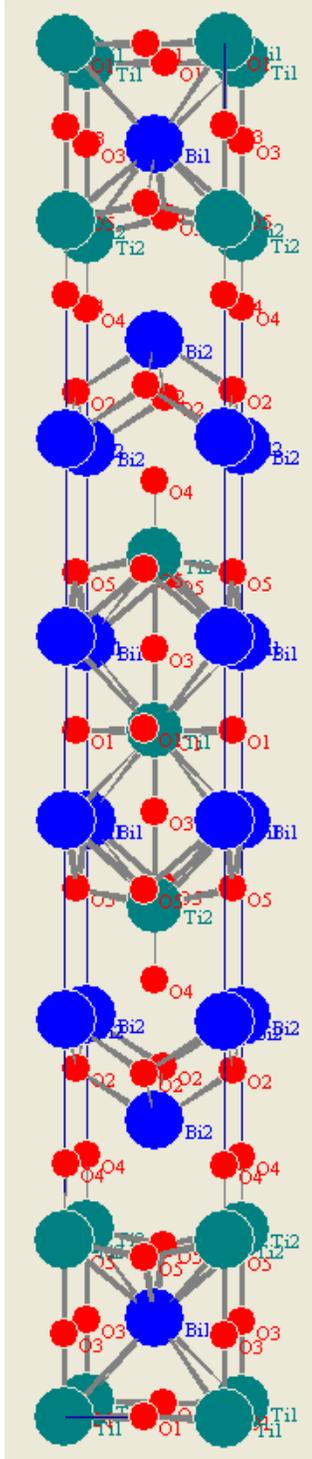

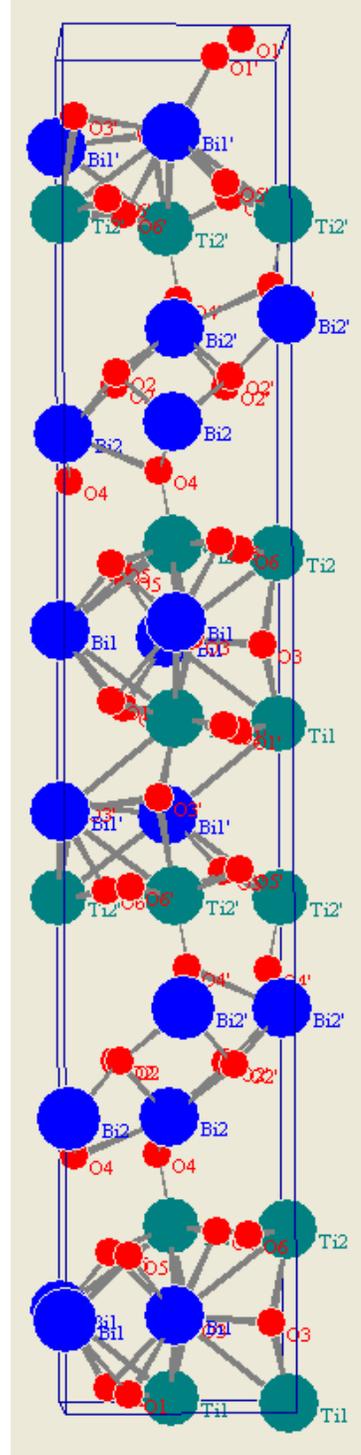

**Fig. 1 (a)**                    **Fig. 1 (b)**

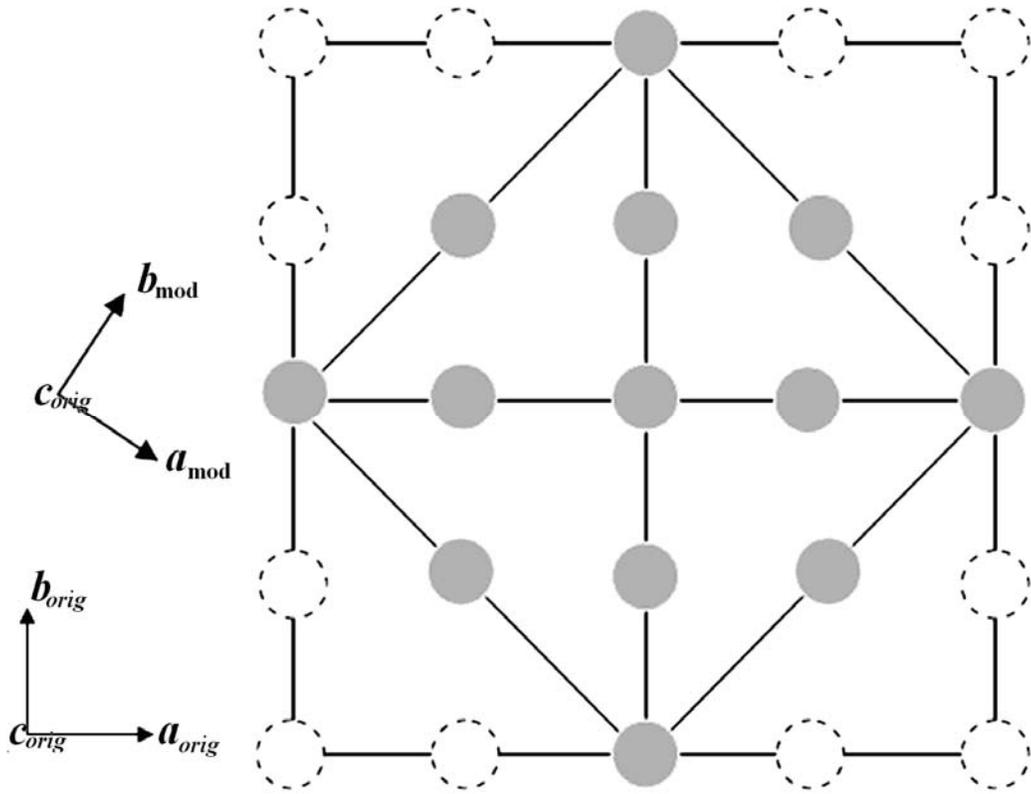

**Fig. 2**

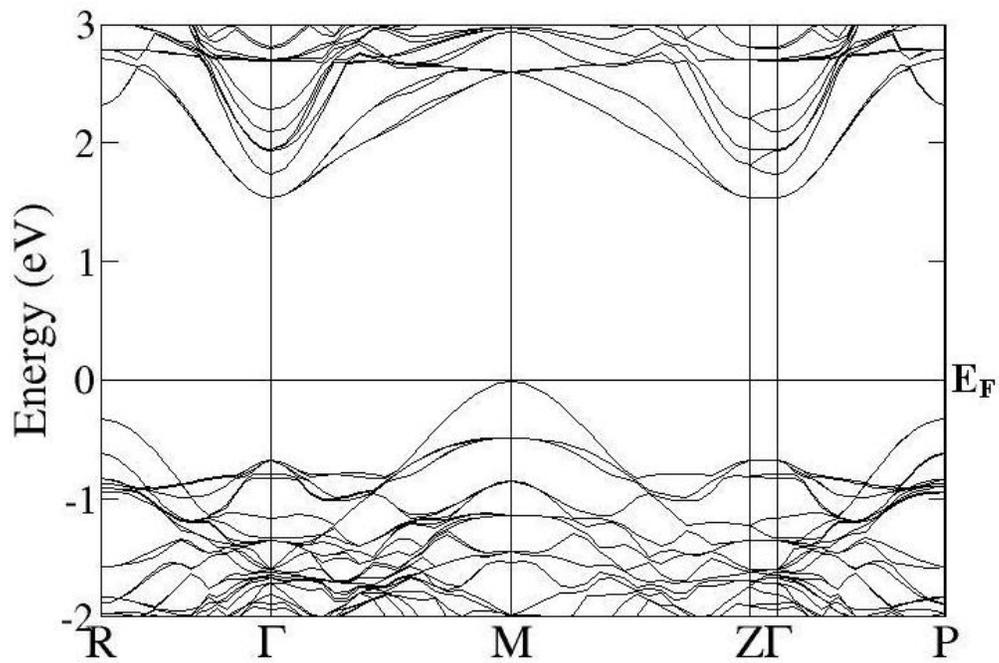

**Fig. 3 (a)**

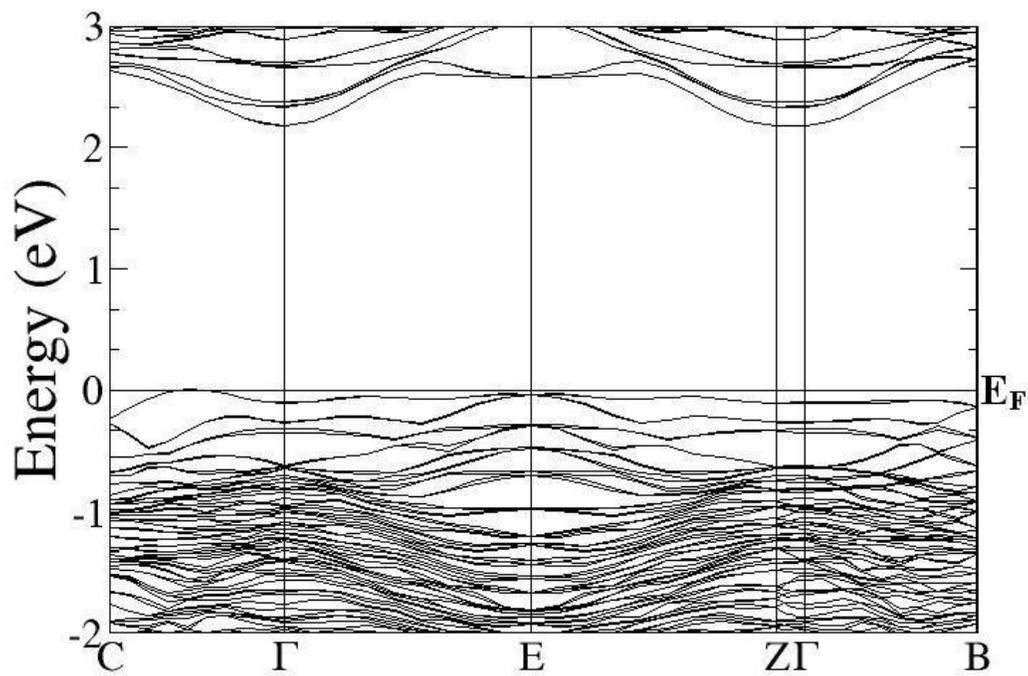

**Fig. 3 (b)**

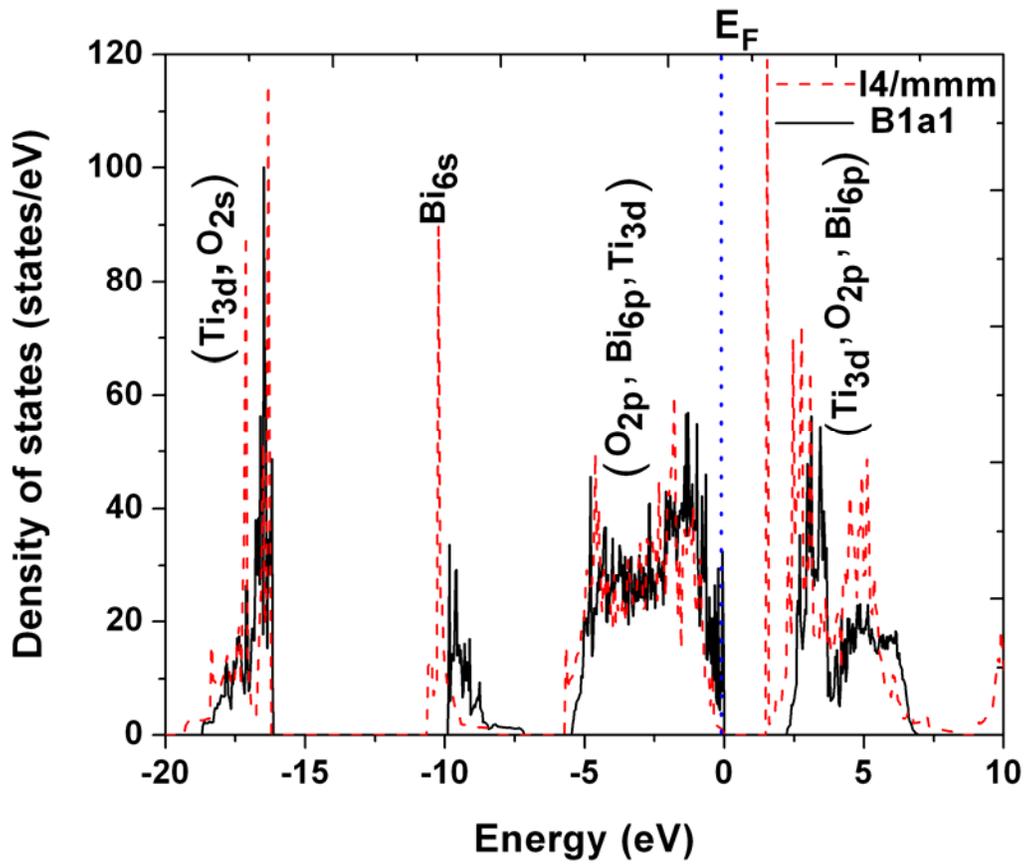

**Fig. 4**

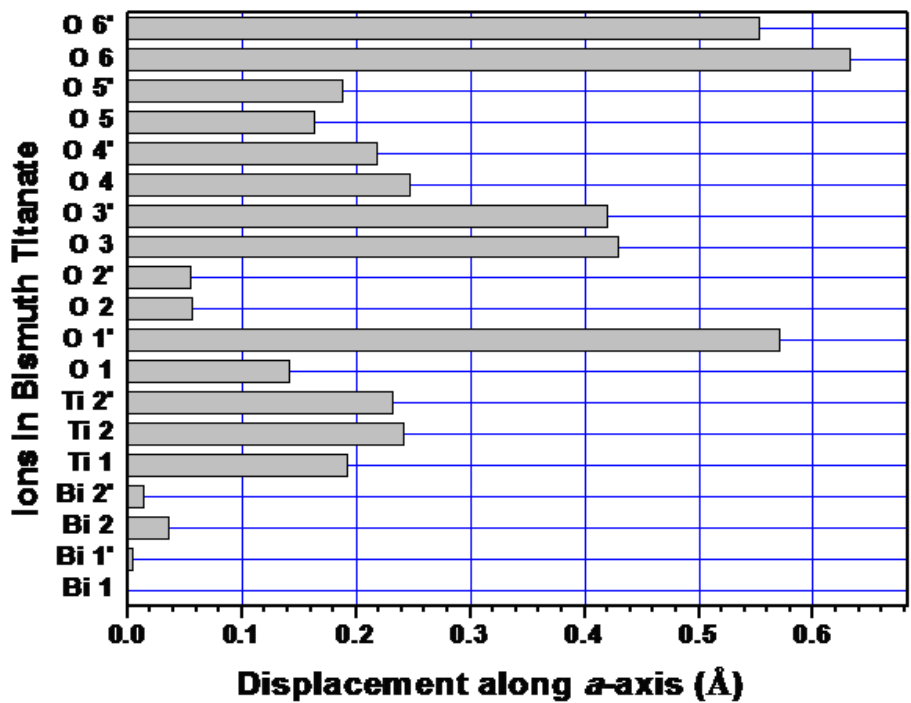

**Fig. 5**